\documentclass[%
 twocolumn,
superscriptaddress,
 amsmath,amssymb,
 aps,
]{revtex4-1}

\usepackage{graphicx}
\usepackage{dcolumn}
\usepackage{bm}
\usepackage{verbatim}
\usepackage{xcolor}
\usepackage{multirow}
\usepackage[flushleft]{threeparttable}
\newcolumntype{.}{D{.}{.}{-1}}
\newcolumntype{d}[1]{D{.}{.}{#1}}
\newcommand*{\wn}{cm$^{-1}$}

\begin{document}

\title{Fourier-transform VUV spectroscopy of $^{14,15}$N and $^{12,13}$C}

\author{K.-F. Lai}
 \affiliation{Department of Physics and Astronomy, LaserLaB, Vrije Universiteit \\
 De Boelelaan 1081, 1081 HV Amsterdam, The Netherlands}
\author{W. Ubachs}%
 \affiliation{Department of Physics and Astronomy, LaserLaB, Vrije Universiteit \\
 De Boelelaan 1081, 1081 HV Amsterdam, The Netherlands}
 \author{N. de Oliveira}%
  \affiliation{Synchrotron Soleil, Orme des Merisiers, St. Aubin, BP 48, 91192 Gif sur Yvette Cedex, France}
 \author{E. J. Salumbides}%
  \affiliation{Department of Physics and Astronomy, LaserLaB, Vrije Universiteit \\
 De Boelelaan 1081, 1081 HV Amsterdam, The Netherlands}

\date{\today}

\begin{abstract}

Accurate Fourier-transform spectroscopic absorption measurements of vacuum ultraviolet transitions in atomic nitrogen and carbon were performed at the Soleil synchrotron.
For $^{14}$N transitions from the $2s^22p^3\,^4$S$_{3/2}$ ground state and  from the $2s^22p^3\,^2$P and $^2$D metastable states were determined in the $95 - 124$ nm range at an accuracy of $0.025\,\mathrm{cm}^{-1}$. Combination of these results with data from previous precision laser experiments in the vacuum ultraviolet range reveal an overall and consistent offset of -0.04 \wn\ from values reported in the NIST database.
The splittings of the $2s^22p^3\,^4$S$_{3/2}$ --
$2s2p^4\,^4$P$_{J}$
transitions are well-resolved for $^{14}$N and $^{15}$N and isotope shifts determined. While excitation of a $2p$ valence electron yields very small isotope shifts, excitation of a $2s$ core electron results in large isotope shifts, in agreement with theoretical predictions.
For carbon six transitions from the ground $2s^22p^2\,^3$P$_{J}$ and $2s^22p3s\, ^3$P$_{J}$ excited states at $165$ nm are measured for both $^{12}$C and $^{13}$C isotopes.

\end{abstract}

\maketitle

\section{Introduction}
The determination of level energies in first row atoms critically relies on accurate spectroscopic measurements in the vacuum ultraviolet (VUV) region below the atmospheric absorption cutoff. The present study applies a unique Fourier-transform spectroscopic instrument in combination with synchrotron radiation to access this wavelength range at high resolution and accuracy for improving the atomic level structures of N and C atoms, including isotopic effects.

The currently available level energies and line classifications for the N atom, compiled in the comprehensive NIST database~\cite{NIST_ASD}, mostly originate from the work of Eriksson and coworkers from late 1950s in combination with the work by Kaufman and Ward~\cite{Kaufman1967}. Eriksson measured N I transitions between 113.4 - 174.5 nm at about 0.1 \wn\ accuracy and constructed the atomic level structure, also including information on transitions between excited states in the visible and IR region \cite{Eriksson1958,Eriksson1961}.
Kaufman and Ward measured the $2p^3$ $^2$D$_{J}$ - $3s$ $^2$P$_J$ and $2p^3$ $^2$P$_{J}$ - $3s$ $^2$P$_J$ transitions to extend the knowledge of the level structure of the ground configuration at better than 0.04 \wn\ accuracy~\cite{Kaufman1967}, also including the forbidden transition $^4$S$_{1/2}$ - $^2$P$_J$ measured by Eriksson. Further analyses were performed by Eriksson~\cite{Eriksson1971,Eriksson1974}, and a compilation was made by Moore~\cite{Moore1975}, now used as a primary reference in the NIST database. Eriksson published an extensive analysis with newly determined energy levels
at an uncertainty of 0.003 \wn~\cite{Eriksson1986}. More recently, Salumbides et al.~\cite{Salumbides2005} measured 12 transitions from the ground state at around 96 nm using VUV precision laser spectroscopy with 0.005 \wn\ uncertainty, thus providing an accurate connection between the ground and excited states.

The energy level structure and the spectral data for the neutral carbon atom were recently reviewed by Haris and Kramida~\cite{Haris2017}. Their report includes an accurate summary of the VUV transitions in C I~\cite{Haris2017}. Among the body of reported studies, the VUV measurements by Kaufman and Ward~\cite{Kaufman1966} present the highest accuracy, at 0.025 - 0.047 \wn, in the range 145 - 193 nm. The C I level energy optimization also includes accurate unpublished UV Fourier-transform data at about 0.004 \wn\ accuracy from Griesmann and Kling reported in Ref.~\cite{Haris2017}. Moreover, accurate VUV laser measurements at 94 nm~\cite{Labazan2005}, allow for the determination of some key excited level energies, accurate to 0.0013 \wn.

Atomic isotope shifts (IS) have been studied for a variety of transitions in $^{12,13}$C. Yamamoto et al.~\cite{Yamamoto1991} and Klein et al. ~\cite{Klein1998} have performed high precision IS measurements of the far-infrared lines of the $^3$P ground term. For transitions between electronic states anomalous, negative IS have been measured, e.g. for the $2p^2$ $^1$S$_0$ - $3s$ $^1$P$_1$ transition~\cite{Burnett1950,Holmes1951}, while the  transition between ground $^3$P$_2$ and core-excited state $2s2p^3$ $^5$S$_2$ showed a positive IS, with the heavier isotope blue-shifted~\cite{Bernheim1980}. Ground state excitation to the autoionizing $2s2p^3$ $^3$S$_1$ state also yielded a positive IS~\cite{Labazan2005}, while on the other hand excitation to $2s2p^3$ $^3$D$_J$ and  $3s$ $^3$P$_J$, as well as the $^1$D$_2$ - $3s$ $^1$P$_1$ transition exhibit an IS with opposite sign~\cite{Haridass1994}. Berengut et al.~\cite{Berengut2006} performed theoretical studies on isotopic shifts of C I, explaining  significant differences in IS for various transitions.

The IS of the nitrogen atom was investigated as well, but mainly in excitation between excited states. Holmes studied the $^{14,15}$N IS of  lines around 800 nm by classical means, finding a  -0.4 to -0.6 \wn\ IS for the $3s$ $^4$P - $3p$ $^4$L quartet transitions and about 0.07 \wn\ for $3s$ $^2$P - $3p$ $^2$P doublet transitions~\cite{Holmes1943,Holmes1951}. Later a number of Doppler-free laser saturation studies were performed on the $3s$ $^4$P - $3p$ $^4$P,$^4$D  transitions~\cite{Cangiano1994,Jennerich2006}. A strong $J$-dependence of the specific mass shift (SMS) effect
was found to originate from the lower $3s$ $^4$P state~\cite{Carette2010}.
The only measurement of IS in VUV transitions from the $2p^3$ $^4$S$_{3/2}$ ground state is that of Salumbides et al.~\cite{Salumbides2005} probing $4s$ $^4$P, $3d$ $^2$F, $3d$ $^4$P, $3d$ $^4$D and $3d$ $^2$D states, 
where no significant $J$-dependent SMS was observed.

In the present study lines of N I in the range of 95 - 124 nm and C I at 165 nm are investigated by Fourier-transform synchrotron absorption spectroscopy with accurate determination of isotopic shifts. Some 27 lines of N I and six lines of the $2p^2$ $^3$P$_J$ - $3s$ $^3$P$_J$ multiplet of C I are measured at an absolute accuracy of 0.025 \wn.

\section{Experiment}

The vacuum ultraviolet (VUV) Fourier-Transform (FT) spectroscopic instrument at the DESIRS beamline of the Soleil Synchrotron facility has been described in detail previously~\cite{Deoliveira2011,Deoliveira2016}. The main originality of the instrument lies in the use of wave-front division interferomtery for generating an interferogram avoiding transmission optical materials for beam overlap. As is common in non-dispersive FT spectroscopy, the absorbing gas cell is located between the source and the VUV interferometer,  allowing for a geometry where the absorption may occur far-distanced from the FT-analyzing instrument.

\begin{figure}[b]
\begin{center}
\includegraphics[width=0.7\linewidth]{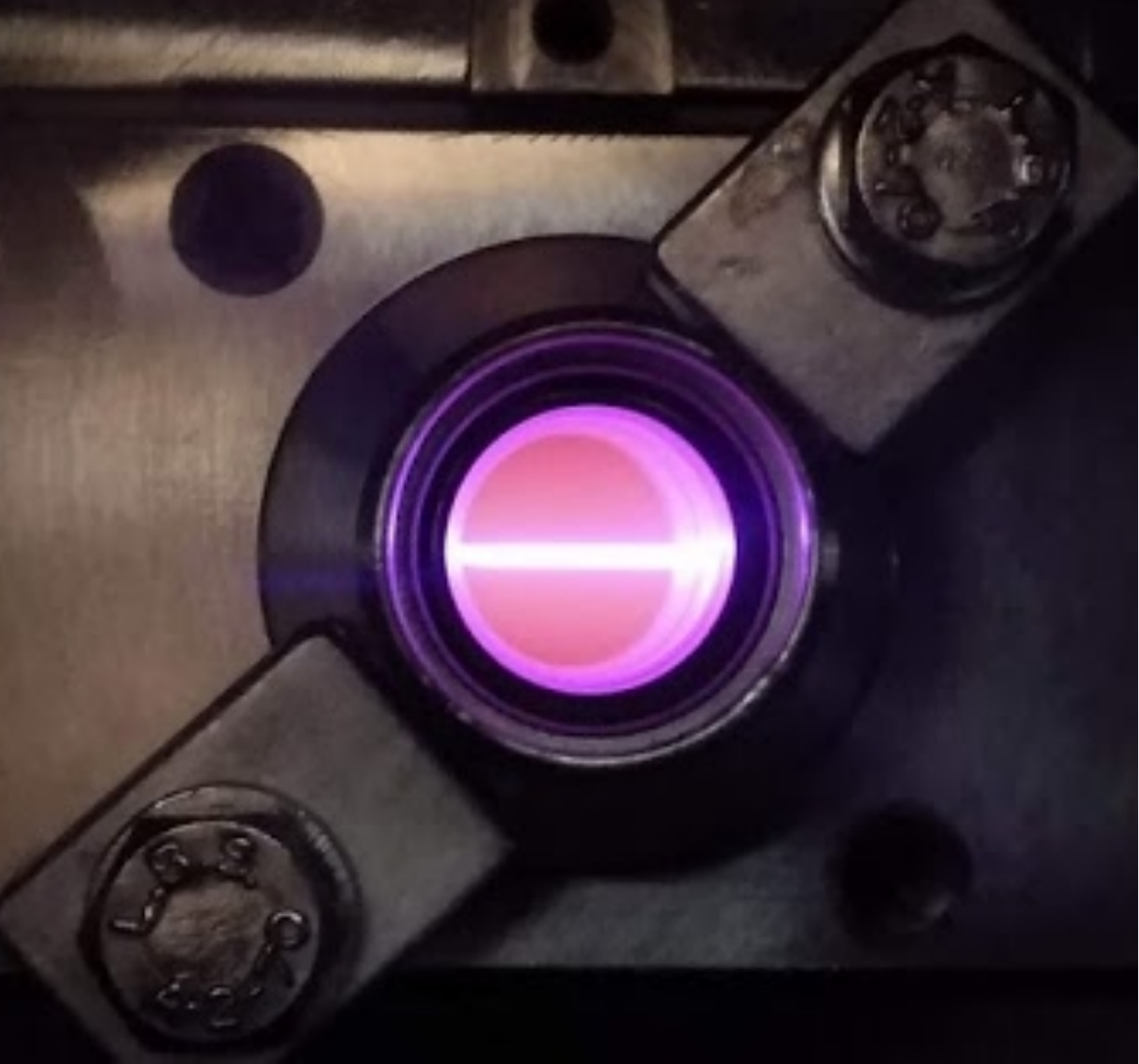}
\caption{\label{Plasma}
Nitrogen plasma glow in the gas filter chamber of the DESIRS beam line with the synchrotron pencil beam traversing. }
\end{center}
\end{figure}

The energy calibration of the FT spectrometer is intrinsically related to the VUV optical path difference which is measured via interferometric detection of fringes of a HeNe laser probing the back surface of the moving reflector~\cite{Deoliveira2011}. The FT spectroscopy energy scale is strictly linear and in principle requires a single reference if precise absolute calibration is needed. Such calibration of the spectra is performed by comparing with an absorption line of Kr I present in most spectra, for which an accurate value exists in the literature at 85846.7055 (2) \wn~\cite{Sansonetti2007}. This value for a natural sample of krypton is in agreement with isotope-specific calibrations by high-resolution laser measurements~\cite{Trickl2007}. This leads to an estimated uncertainty of 0.025 \wn\ for the N I and C I resonances. The widths of observed lines range from 0.28 to 0.40 \wn~(FWHM).

In a variety of previous experiments the FT-instrument was applied to perform spectroscopy of quasi-static gas molecules flowing from a capillary-shaped windowless gas cell, such as for nitrogen molecules~\cite{Heays2019}. Recently a study was performed probing Rydberg states of O$_2$ molecules in excitation from metastable states down to 120 nm, produced in a DC-discharge cell equipped with UV-transmissive windows~\cite{Western2020}.
In the present study spectroscopy  is performed on atoms that were produced via two entirely different methods.

The measurements on atomic nitrogen were carried out by releasing N$_2$-gas into a windowless gas filter, located upstream on the DESIRS beam line close to the undulator. This filter is usually filled with noble gas for the purpose of suppressing the high energy harmonics produced in the undulator~\cite{Mercier2000}. The gas density, nor the absorbing column length, are known, but the gas inlet can be controlled to produce the desired signal strength, where a strong limitation is set by the maximum pressure allowed before the safety shutters on the beam line close. The gas filter can be monitored through a viewing window where a radiation emitting plasma can be observed of blue-purple color at the location where the synchrotron pencil beam traverses (see Fig.~\ref{Plasma}). At this location the synchrotron beam, including its harmonics produced in the undulator, causes photo-dissociation and photo-ionization in a collisional environment, and hence a plasma, where N atoms are produced in the ground state as well as in the metastable states. The absorption spectra of this nitrogen plasma are measured by the FT-instrument some 17 m further downstream. During the measurements gas samples of $^{14}$N$_2$, $^{14}$N$^{15}$N, and $^{15}$N$_2$ were used to measure and disentangle the isotopic lines of N I.

For the spectroscopy of atomic carbon, a DC discharge cell is used, which is located further downstream just in front (by 0.5 m) of the FT-instrument inside the conventional gas sample chamber of the FT spectroscopy branch~\cite{Deoliveira2016}. The DC-discharge is similar to the one used by Western et al.~\cite{Western2020}, although the cell is windowless in this case in order to reach the VUV spectral range. A flow of CO$_2$ gas is released at the inlet port, and pumped at the rear end. A plasma is generated between the cathodes at a Voltage difference of 1000 V with a stabilized discharge current of 20 mA. The discharge is further stabilized by optimizing the pressure and by mixing in of He carrier gas, within the limits allowed by the differential pumping system of the chamber~\cite{Deoliveira2016}.
Spectra are recorded for $^{12}$C and $^{13}$C by using $^{12}$CO$_2$ and enriched $^{13}$CO$_2$ gas.

\section{Results and Interpretation}

\subsection{Nitrogen I}

Two different sets of lines are measured for N I, lines in excitation from the $^4$S$_{3/2}$ ground state and lines excited from $^2$D$_J$ metastable states produced in the plasma. Results will be discussed separately.

\subsubsection{Initial state: Ground $2s^22p^3\,^4$S$_{3/2}$}

\begin{figure}[t]
\begin{center}
\includegraphics[width=\linewidth]{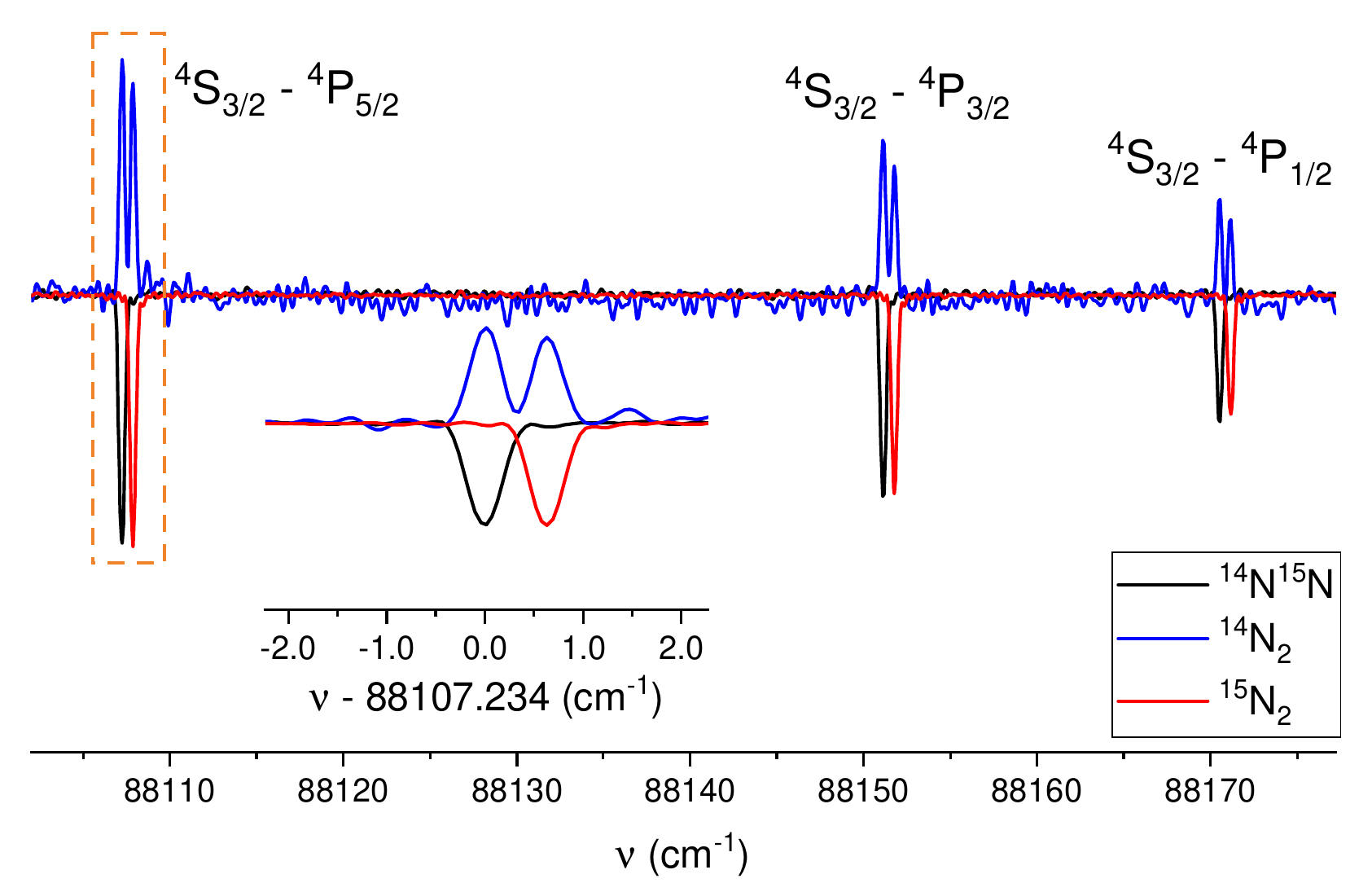}
\caption{\label{N_2p4}
Overview spectra of N I core changing transitions $^4$S$_{3/2}$ - $2s2p^4$ $^4$P$_{J}$ excited from the ground state, using different isotopic parent gases. The inset presents the $^4$S$_{3/2}$ - $2s2p^4$ $^4$P$_{5/2}$ line exhibiting a well-resolved isotopic shift.}
\end{center}
\end{figure}

Recorded absorption spectra from the $2s^22p^3\,^4$S$_{3/2}$ ground state to the $3s$ $^4$P$_{J}$ levels and the $2s2p^4$ $^4$P$_{J}$ core excited levels were measured, the latter shown in Fig.~\ref{N_2p4}.
Spectra were recorded from samples of $^{14}$N$_2$, $^{14}$N$^{15}$N, and $^{15}$N$_2$, thus allowing for unraveling the isotopic structure.
Table \ref{tab:ground N} lists the transition frequencies as deduced from the spectra. For the core excited states a clear isotopic splitting was observed, results of which are included in the Table.
The same lines of $^{14}$N I are well studied by Kaufman and Ward~\cite{Kaufman1967} with uncertainties of 0.06 to 0.1 \wn. When comparing the present data set with that of Ref.~\cite{Kaufman1967} an averaged systematic offset of -0.05 \wn\ is found, corresponding to  1.5 $\sigma$ of combined uncertainties.

The agreement between the present FT-data with the previous VUV-laser data in the range above 104,000 \wn~\cite{Salumbides2005}, except for the line exciting the $^4$P$_{1/2}$ level (off by $2\sigma$), is considered as a verification of the calibration accuracy of the present experiment.

\begin{table}
\renewcommand{\arraystretch}{1.3}
\caption{\label{tab:ground N}
Measured transition frequencies for N I and isotopic shifts for lines excited from the $2s^22p^3\,^4$S$_{3/2}$ ground state.
Units in \wn\ with uncertainties indicated in parentheses. In the fourth column the derived transition frequencies of Ref.~\cite{Kaufman1967} are listed for comparison. For the transitions above 100,000 \wn\ a comparison is made with results of the precision VUV laser study~\cite{Salumbides2005}. The isotopic shift ($^{15}$N-$^{14}$N), $\Delta_{15-14}$, is given in last column. In cases where no significant isotope shift is observed the value in parentheses represents an upper limit.}
\begin{tabular}{ccccc}
Excited state &  $J$	& \multicolumn{1}{c}{$^{14}$N}& \multicolumn{1}{c}{Ref. \cite{Kaufman1967,Salumbides2005}} & \multicolumn{1}{c}{$\Delta_{15-14}$}\\
\hline
& 1/2  &83284.021\,(25)&83284.085(42)& 0.01(4)\\
$3s$ $^4$P & 3/2  &83317.784\,(25)&83317.843(42)&0.00(4)\\
& 5/2  &83364.570\,(25)&83364.637(42)&0.01(4)\\
\hline
& 5/2  &88107.226\,(25)&88107.272(39)& 0.64(1)\\
$2s2p^4$ $^4$P & 3/2  &88151.130\,(25)&88151.185(39)& 0.63(1) \\
& 1/2  &88170.525\,(25)&88170.585(39)& 0.63(1)\\
\hline
& 5/2  &104825.080\,(25)&104825.0699(50)& 0.00(4)\\
$3d$ $^4$P & 3/2  &104859.698\,(25)&104859.6952(50)& 0.00(4) \\
& 1/2  &104886.012\,(25)&104886.0687(50)& 0.02(4)\\
\hline
$3d$ $^2$F & 5/2  &104810.335\,(25)&104810.3324(50)& 0.01(4) \\
\hline
\end{tabular}
\end{table}

The $^4$S$_{3/2}$ - $3s$ $^4$P$_{J}$ transition frequencies show no significant difference in measurement using pure $^{14}$N$_2$ and $^{15}$N$_2$, while from measurements with $^{14}$N$^{15}$N the spectra show insignificant change in linewidth. The uncertainty of FWHM is about 0.021 \wn, which is dominated by the statistics and the deconvolution of the FT sinc shape of the apparatus function observed in the FT spectrum. From this it is estimated that the $^{15}$N - $^{14}$N (IS) of the $^4$S$_{3/2}$ - $3s$ $^4$P$_{J}$ transitions is less than 0.04 \wn. The transitions exciting $3d$ states also do not display an isotope shift. This is found to be in agreement with the previous, more accurate, VUV laser experiment where an IS of 0.01 \wn\ was determined~\cite{Salumbides2005}.

In contrast, the $^4$S$_{3/2}$ - $2s2p^4$ $^4$P$_{J}$ core changing transitions display a distinctive IS as clearly shown in Fig.~\ref{N_2p4}. The spectra of $^{14}$N$^{15}$N show a well resolved isotopic doublet for each individual fine structure line with nearly equal intensity. The transition frequencies are consistent with those found from pure $^{14}$N$_2$ and $^{15}$N$_2$. The $^{15}$N - $^{14}$N IS are determined to be about 0.63(1) \wn\ with an uncertainty determined by line profile fitting in the $^{14}$N$^{15}$N scan.

\subsubsection{Initial state: Metastable $2s^22p^3\,^2$D$_{J}$}

\begin{figure}
\begin{center}
\includegraphics[width=\linewidth]{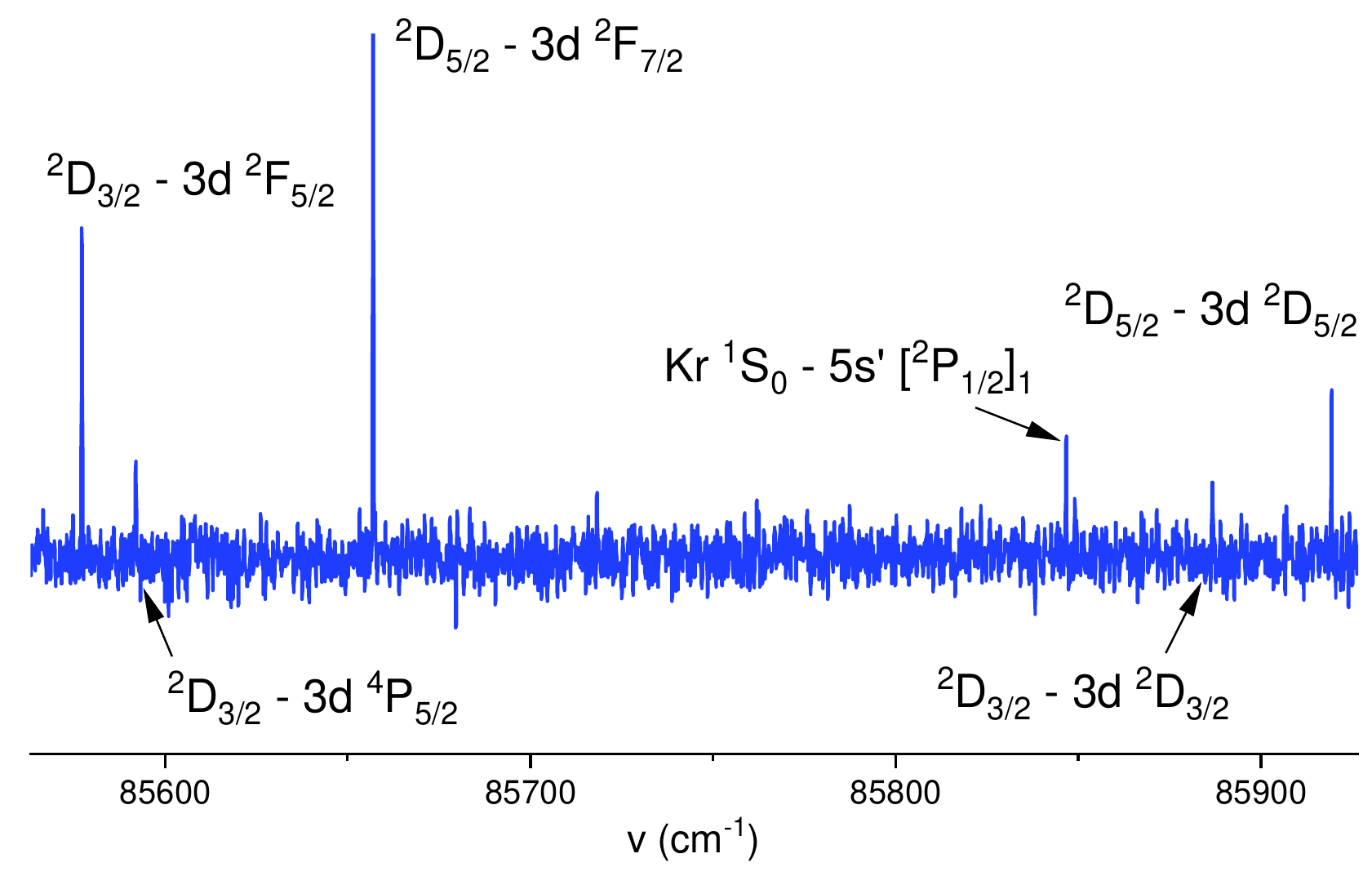}
\caption{\label{N_meta_overview}
Spectra of $^{14}$N I transitions excited from the metastable $^2$D$_J$ states in the range 85500 - 86000 \wn. Also shown is the reference Kr I used for calibrationo}
\end{center}
\end{figure}

The metastable $2s^22p^3$ $^2$D$_{J}$ states lie some 2.5 eV above the ground state $^4$S$_{3/2}$, 19224.464 \wn\ for $^2$D$_{5/2}$ and 19233.177 \wn\ for $^2$D$_{3/2}$~\cite{NIST_ASD}. All transitions excited from these levels are substantially weaker than those excited from the $^4$S$_{3/2}$ ground state, indicating that the metastable states are less populated.
A spectrum if these lines is shown in Fig.~\ref{N_meta_overview}, while Table \ref{tab:meta N} lists all the observed transitions and their frequencies.
Regardless of the small splitting, of about 0.485 \wn\ between $3s$ $^2$D$_{5/2}$ and $3s$ $^2$D$_{3/2}$, the transition observed at 80430.79 \wn\ and 80439.02 \wn\ are unambiguously assigned with $^2$D$_{3/2}$ - $3s$ $^2$D$_{3/2}$ and $^2$D$_{5/2}$ - $3s$ $^2$D$_{5/2}$, based on intensity.
Note that $\Delta J = 0$ transitions of $^2$D$_{J'}$ - $3s$ $^2$D$_{J}$ exhibit Einstein $A$ coefficients 10-times larger than $\Delta J = \pm1$.
Table~\ref{tab:meta N} includes the transition frequencies of observed lines from Ref.~\cite{Kaufman1967} for comparison. Here again a systematic offset is found, now positive and of opposite sign at 0.078 \wn, again corresponding to 1.5 $\sigma$ of combined uncertainties. The same systematic offset of -0.07~\wn\ is also found when comparing the results of the VUV laser measurements~\cite{Salumbides2005} with those of Kaufman~\cite{Kaufman1967}.

\begin{table}
\renewcommand{\arraystretch}{1.3}
\begin{threeparttable}
\caption{\label{tab:meta N}
Measured transition frequencies for lines excited from $^2$D$_J$ and $^2$P$_J$ metastable states in N I, with uncertainties indicated in parentheses. Again a comparison is made with results from Ref.~\cite{Kaufman1967}. All values  in \wn.}
\begin{tabular}{cccc}
Initial state	& Excited state	& \multicolumn{1}{c}{$^{14}$N}& Ref. \cite{Kaufman1967}  \\
\hline
& $3s$ $^{2}$D$_{5/2}$  &80439.004\,(25) &80438.900(78)\\
& $4s$ $^{2}$P$_{3/2}$  &84997.212(25) &84997.174(79)\\
& $3d$ $^{2}$F$_{7/2}$  &85656.939\,(25) &85656.890(81)\\
$^{2}$D$_{5/2}$ & $3d$ $^{2}$D$_{5/2}$  &85919.307\,(25) & 85919.252(81)\\
& $5s$ $^{2}$P$_{3/2}$  &90879.420\,(25) & 90879.406(83)\\
& $4d$ $^{2}$F$_{7/2}$  &91138.025\,(25) & 91138.01(17)$^a$\\
& $5d$ $^{2}$F$_{7/2}$  &93666.807\,(25) & 93666.64(18)$^a$\\
\hline
& $3s$ $^{2}$D$_{3/2}$  &80430.769\,(25) & 80430.684(71)\\
& $4s$ $^{2}$P$_{1/2}$  &84911.672\,(25) &84911.643(72)\\
& $3d$ $^{2}$F$_{5/2}$  &85577.230(25) &85577.181(73)\\
$^{2}$D$_{3/2}$& $3d$ $^{4}$P$_{5/2}$  &85591.960\,(25) &85591.933(73)\\
& $3d$ $^{2}$D$_{3/2}$  &85886.730\,(25) &85886.702(66)\\
& $5s$ $^{2}$P$_{1/2}$  &90802.570\,(25) & 90802.522(66)\\
& $4d$ $^{2}$F$_{5/2}$  &91053.168\,(25) & 91053.04(17)$^a$\\
& $4d$ $^{4}$P$_{5/2}$  &91066.829\,(25) & 91066.64(17)$^a$\\
\hline
$^{2}$P$_{3/2}$& \multirow{2}{*}{($^1$S) $3s$ $^2$S$_{1/2}$}  &87439.234(25) &87439.252$^b$\\
$^{2}$P$_{1/2}$&  &87439.717(25) &87439.638$^b$\\
\hline
\end{tabular}
\begin{tablenotes}
\footnotesize
\item $^a$Measured frequency presented in Ref.
\cite{Eriksson1971}.
\item $^b$Calculated frequency from NIST database.
\end{tablenotes}
\end{threeparttable}
\end{table}

Two additional transitions are observed near 87439 \wn, which we  assign as $2p^3$ $^2$P$_{1/2}$,  $^2$P$_{3/2}$ - ($^1$S) $3s$ $^2$S$_{1/2}$. Eriksson first identified a line at 114.3649(2) nm, or 87439.42(15) \wn~\cite{Eriksson1958}.
From the level energies reported in the NIST database frequencies for these two transitions can be computed, and these predictive results are found to be well in agreement with the present direct measurements (see Table \ref{tab:meta N}).
The fine-structure splitting of $2p^3$ $^2$P$_{3/2}$ - $^2$P$_{1/2}$ is determined to be 0.483(35) \wn, which is in fair agreement with the paramagnetic resonance result of 0.4326(3) \wn~\cite{Diebold1982}.

The transitions from metastable states show no significant IS from spectra obtained with $^{14}$N$_2$ and $^{15}$N$_2$ gases. The linewidth obtained with $^{14}$N$^{15}$N gas does not show any additional broadening, from which it is concluded that the IS of these lines is smaller than $0.04$ \wn.

\subsection{Carbon I}

\begin{figure}[t]
\begin{center}
\includegraphics[width=\linewidth]{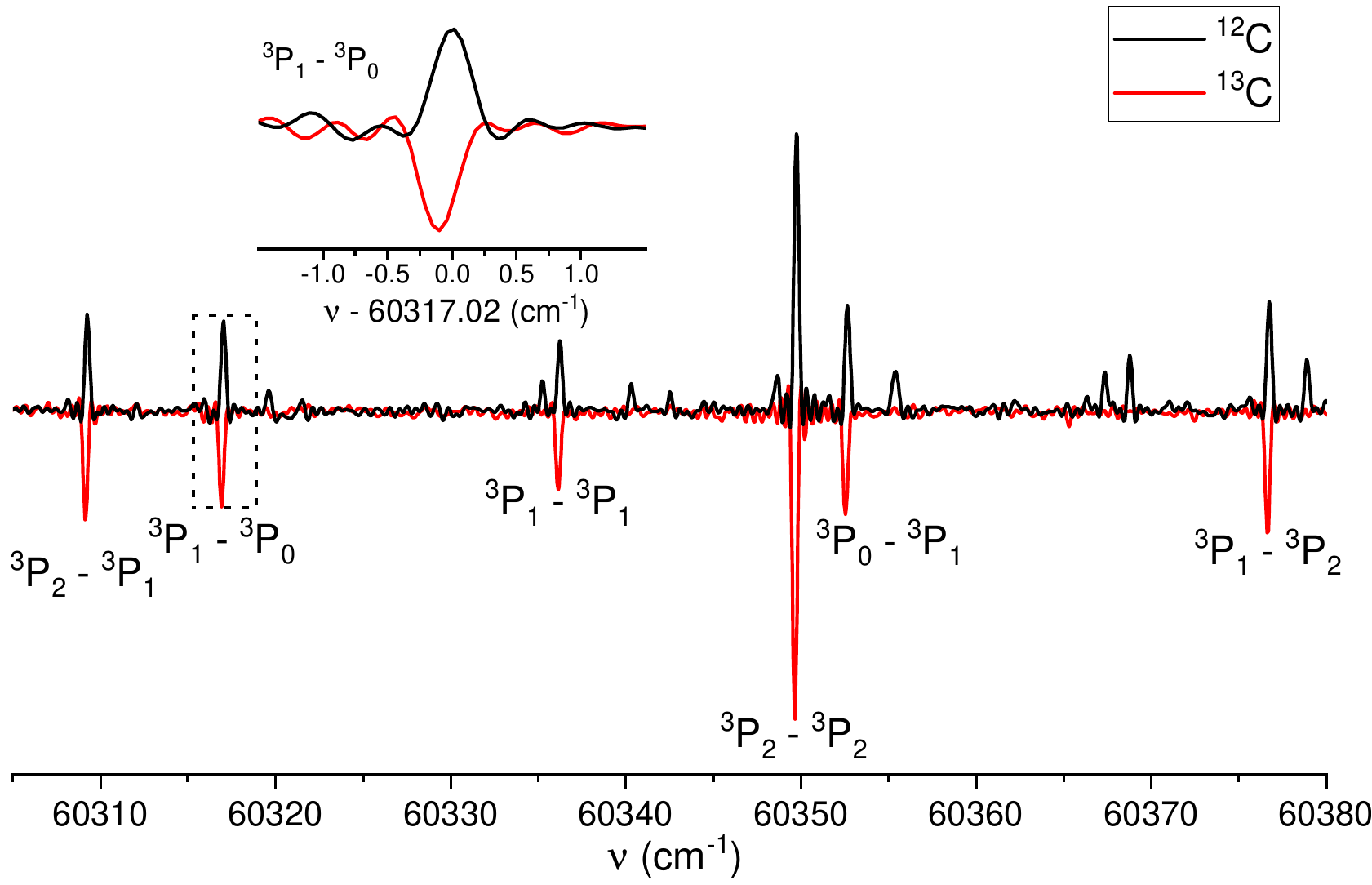}
\caption{\label{C_3s}
Overview spectra of C I $^3$P$_J$ - $3s$ $^3$P$_J$ multiplet. The inset shows the $^3$P$_1$ - $3s$ $^3$P$_0$ transition. The $^{12}$C spectrum is partly overlapped with lines from the A-X(2,0) band of $^{12}$C$^{16}$O.
}
\end{center}
\end{figure}

Figure~\ref{C_3s} shows the recording of all six $2p^2$ $^3$P$_J$ - $3s$ $^3$P$_J$ transitions for both carbon isotopes obtained from discharges in $^{12}$CO$_2$/He and $^{13}$CO$_2$/He  gas mixtures. The spectrum of $^{12}$C has several overlapping lines of the A$^1\Pi$-X$^1\Sigma^+$ (2,0) band of $^{12}$CO~\cite{Niu2016}.
The A-X(2,0) band of $^{13}$C$^{16}$O is blue-shifted by 100 \wn, outside the measurement interval displayed. The spectra of $^{12}$C and $^{13}$C are taken in separate measurements in absence of the Kr I reference line in the scan range.

The absolute calibration is verified by interpolating from the $^{12}$CO lines with predicted transition frequencies from Doppler-free two-photon spectroscopy results at 0.002 \wn\ accuracy~\cite{Niu2013,Niu2016}. The averaged difference of the twelve CO lines observed is 0.001(6) \wn, which is smaller than the current measurement uncertainty. This procedure again leads to the same accuracy as for the N I lines, at 0.025 \wn.

\begin{table*}
\renewcommand{\arraystretch}{1.3}
\begin{threeparttable}
\caption{\label{tab:C}
Measured frequency and isotopic shift ($\Delta_{13-12}$) of C I $^3$P$_J$ - $3s ^3$P$_J$ transitions, with uncertainties indicated in parentheses. Note that the  $^{12}$C transition frequencies from Ref. \cite{Haris2017} are computed values. The uncertainty of isotopic shifts from Ref. \cite{Haridass1994} is estimated by taking measurement uncertainties from $^{12}$C and $^{13}$C in quadrature. All values given in \wn.}
\begin{tabular}{cd{8}d{8}d{8}d{9}d{9}d{7}.}
\hline
\hline
& \multicolumn{3}{c}{This work}	& \multicolumn{2}{c}{Ref. \cite{Haris2017}}&\multicolumn{2}{c}{Ref. \cite{Haridass1994}}\\
Transition	& \multicolumn{1}{c}{$^{12}$C}	& \multicolumn{1}{c}{$^{13}$C} & \multicolumn{1}{c}{$\Delta_{13-12}$} &\multicolumn{1}{c}{$^{12}$C}	&  \multicolumn{1}{c}{$\Delta_{13-12}$}&\multicolumn{1}{c}{$^{12}$C}	&  \multicolumn{1}{c}{$\Delta_{13-12}$}\\
\hline
$^{3}$P$_{2}$ - $3s^{3}$P$_{1}$&60309.245(25) & 60309.141(25)& -0.104(35)&  60309.2459\,(13) & -0.0919(28)&  60309.22(10) & \multicolumn{1}{c}{-} \\
$^{3}$P$_{1}$ - $\quad^{3}$P$_{0}$&17.021(25) &16.919(25) & -0.102(35)&  17.0319\,(14)  & -0.0919(28)&  17.02(10)  & -0.3 ^a\\
$^{3}$P$_{1}$ - $\quad^{3}$P$_{1}$&36.234(25)&36.148(25) & -0.086(35)&   36.2427\,(13) & -0.0917(28)&   36.22(10) & -0.15\\
$^{3}$P$_{2}$ - $\quad^{3}$P$_{2}$&49.750(25)&49.653(25) & -0.097(35)&   49.7568\,(13) & -0.0918(28)&   49.73(10) & -0.14\\
$^{3}$P$_{0}$ - $\quad^{3}$P$_{1}$&52.663(25)&52.549(25) & -0.114(35)&   52.6594\,(13) & -0.0914(28)&   52.64(10) & -0.14\\
$^{3}$P$_{1}$ - $\quad^{3}$P$_{2}$&76.750(25)&76.649(25) & -0.101(35)&   76.7536\,(13) & -0.0916(28)&   76.73(10) & -0.16\\
\hline
\hline
\end{tabular}
\begin{tablenotes}
\footnotesize
\item $^{a}$In Ref.~\cite{Haridass1994} it was stated that the corresponding $^{13}$C transition is strongly blended.
\end{tablenotes}
\end{threeparttable}
\end{table*}

\begin{table*}
\renewcommand{\arraystretch}{1.5}
\begin{threeparttable}
\caption{\label{tab:level energy}
Least-squares fitted C I level energies for the $^3$P$_J$ - $3s ^3$P$_J$ multiplet and comparison with Ref. \cite{Haris2017}. The predicted transitions between ground $^3$P$_J$ are compared with high precision measurements in  Ref. \cite{Yamamoto1991} and Ref. \cite{Klein1998}.}
\begin{tabular}{r....}
\hline
\hline
&\multicolumn{2}{c}{$^{12}$C}&\multicolumn{2}{c}{$^{13}$C}\\
	& \multicolumn{1}{c}{This work}	& \multicolumn{1}{c}{Ref. \cite{Haris2017}}& \multicolumn{1}{c}{This work} &\multicolumn{1}{c}{Ref. \cite{Haris2017}}\\
\hline
$2p^2$ $^{3}$P$_{0}$&  0 & 0 & 0 & 0\\
$2p^2$ $^{3}$P$_{1}$&  16.420(30) & 16.4167122(6) & 16.404(30) & 16.4167869(6)\\
$2p^2$ $^{3}$P$_{2}$&  43.419(30) & 43.4134544(8) & 43.405(30) & 43.4136669(16)\\
\hline
$3s\:\:$ $^{3}$P$_{0}$&  60333.441(41) & 60333.4484(14) & 60333.323(41) & 60333.357(4)\\
$3s\:\:$ $^{3}$P$_{1}$&  60352.663(30) & 60352.6594(13) & 60352.549(30) & 60352.568(4)\\
$3s\:\:$ $^{3}$P$_{2}$&  60393.169(40) & 60393.1703(13) & 60393.055(40) & 60393.078(4)\\
\hline
\hline
Transition&\multicolumn{1}{c}{This work}&\multicolumn{1}{c}{Ref. \cite{Yamamoto1991,Klein1998}}&\multicolumn{1}{c}{This work}&\multicolumn{1}{c}{Ref. \cite{Yamamoto1991,Klein1998}}\\
\hline
$^{3}$P$_{0}$ - $^{3}$P$_{1}$& 16.420(30) & 16.4167122(6) & 16.404(30)& 16.416787(1)^a\\
$^{3}$P$_{1}$ - $^{3}$P$_{2}$& 26.999(25) & 26.9967422(6) & 27.001(25)& 26.996881(3)^a\\
\hline
\end{tabular}
\begin{tablenotes}
\footnotesize
\item $^a$Center of gravity for the $^{13}$C hyperfine-resolved transition is reported.
\end{tablenotes}
\end{threeparttable}
\end{table*}

Table~\ref{tab:C} lists the transition frequencies for the six C I lines. The Table includes the transition frequencies of $^{12}$C and the $^{13}$C - $^{12}$C IS. The latter are presented in Ref.~\cite{Haris2017} at better accuracy, but those results do not stem from direct measurement, but rather from combination differences. The results of Haridass et al., directly measured but at larger uncertainty~\cite{Haridass1994}, are included as well. All the measured frequencies agree with each other within the stated uncertainties. Hence the predicted line positions of Ref.~\cite{Haris2017} are confirmed by experiment.

The ground and $3s$ $^3$P$_J$ level energies of $^{12}$C and $^{13}$C are fitted with the six intercombination lines using the LOPT program~\cite{kramida2011}. Table~\ref{tab:level energy} lists the fitted values from this work and a comparison with Ref.~\cite{Haris2017} is made. The level energies and uncertainties presented are relative to the ground $^3$P$_0$ state.
The predicted transition frequencies agree well with high precision measurements of transitions between ground states~\cite{Yamamoto1991,Klein1998}, exhibiting less than 0.012 \wn\ difference.
This finding further supports the calibration accuracy in the present study.
Haris and Kramida~\cite{Haris2017} noted a small systematic shift of -0.00006 nm, or 0.022 \wn, with respect to the $^{12}$C values of Kaufman and Ward~\cite{Kaufman1966} in the $52\,000$ to $78\,000$ \wn\ range of the Griesmann and Kling unpublished FTS data. By comparing these values with our measurement, there appears a 0.019 \wn\ averaged offset, consistent with the claim of Ref.~\cite{Haris2017}.

\section{Discussion}

\subsection{Level energies of $^{14}$N I}

Three transitions from the metastable $2s^22p^3$ $^2$D$_{J}$ state, the transitions $^2$D$_{3/2}$ - $3d$ $^2$D$_{3/2}$, $^2$D$_{5/2}$ - $3d$ $^2$D$_{5/2}$ and $^2$D$_{3/2}$ - $3d$ $^2$F$_{5/2}$, share their excited state in the high precision VUV laser measurements of Ref. \cite{Salumbides2005}. In combination with those laser measurements, the level energies of all states can be fitted using LOPT program~\cite{kramida2011}. Table \ref{tab:N level} lists the level energies of 25 states and makes a comparison with values from the NIST database. A global and consistent negative offset, of -0.04 \wn, is observed. For the $2p^3$ $^2$D$_{J}$ metastable states larger deviations are found of size -0.07 and -0.09 \wn.

The determination of $^2$D$_{J}$ level energies presented in the NIST database is likely based on the VUV measurements by Kaufman and Ward~\cite{Kaufman1967}, which is determined by taking combination differences of transitions $^2$D$_{J}$ - $3s$ $^2$P$_{J}$ and $^2$P$_{J}$ - $3s$ $^2$P$_{J}$ and the forbidden transition $^4$S$_{3/2}$ - $^2$P$_{J}$ measured by Eriksson. The relative uncertainty of the VUV measurements~\cite{Kaufman1967} is tested by comparing the fine-structure splitting of $^2$D$_{J}$ and $^2$P$_{J}$ with results from laser magnetic resonance~\cite{Bley1989} and paramagnetic resonance~\cite{Diebold1982}, respectively. The fitted $^2$P$_{J}$ splitting is 0.391(18) \wn, while the paramagnetic resonance measurement gives 0.4326(27) \wn, reflecting a 2.3 $\sigma$ difference.

The $2s^22p^3\,^2$D$_{5/2}$ - $^2$D$_{3/2}$ splitting extracted from the FTS data here and the VUV-laser study~\cite{Salumbides2005} is 8.72(3) \wn, which is consistent with the accurate value for the fine structure splitting study via laser magnetic resonance yielding 8.720957(7)~\wn~\cite{Bley1989}.

These considerations support additional evidence for the consistency of the present FT-data and the previous VUV-laser precision data. At the same time they support evidence for the inconsistency, i.e. the global shift in the NIST tabulated data for N I levels.

\begin{table}
\renewcommand{\arraystretch}{1.5}
\caption{\label{tab:N level}
Least-squares fitted N I level energies using transition from Ref.~\cite{Salumbides2005} and the current study. The fitted value is compared with the value of the NIST database. All values in \wn.}
\begin{tabular}{r.d{4}.}
Level & \multicolumn{1}{c}{This work \& Ref. \cite{Salumbides2005}} & \multicolumn{1}{c}{NIST} & \multicolumn{1}{c}{Difference}\\
\hline
$2p^3$ $^4$S$_{3/2}$ & \multicolumn{1}{c}{0} & \multicolumn{1}{c}{0} & - \\
$2p^3$ $^2$D$_{5/2}$ & 19224.373(25) & 19224.464 & -0.091\\
$2p^3$ $^2$D$_{3/2}$ & 19233.108(15) & 19233.177 & -0.069\\
$3s$ $^4$P$_{1/2}$ & 83284.021(25) & 83284.070 & -0.049\\
$3s$ $^4$P$_{3/2}$ & 83317.784(25) & 83317.830 & -0.046\\
$3s$ $^4$P$_{5/2}$ & 83364.570(25) & 83364.620 & -0.050\\
$2p^4$ $^4$P$_{5/2}$ & 88107.226(25) & 88107.260 & -0.034\\
$2p^4$ $^4$P$_{3/2}$ & 88151.130(25) & 88151.170 & -0.004\\
$2p^4$ $^4$P$_{1/2}$ & 88170.525(25) & 88170.570 & -0.045\\
$3s$ $^2$D$_{5/2}$ & 99663.377(36) & 99663.427 & -0.050\\
$3s$ $^2$D$_{3/2}$ & 99663.877(29) & 99663.912 & -0.035\\
$4s$ $^4$P$_{1/2}$ & 103622.4773(50) & 103622.51 & -0.03\\
$4s$ $^4$P$_{3/2}$ & 103667.1214(50) & 103667.16 & -0.04\\
$4s$ $^4$P$_{5/2}$ & 103735.4527(50) & 103735.48 & -0.03\\
$4s$ $^2$P$_{1/2}$ & 104144.780(29) & 104144.820 & -0.04\\
$4s$ $^2$P$_{3/2}$ & 104221.585(36) & 104221.630 & -0.045\\
$3d$ $^2$F$_{5/2}$ & 104810.3327(48) & 104810.360 & -0.0273\\
$3d$ $^2$F$_{7/2}$ & 104881.312(36) & 104881.350 & -0.038\\
$3d$ $^4$P$_{5/2}$ & 104825.0702(48) & 104825.110 & -0.040\\
$3d$ $^4$P$_{3/2}$ & 104859.6953(49) & 104859.73 & -0.04\\
$3d$ $^4$P$_{1/2}$ & 104886.0684(49) & 104886.10 & -0.03\\
$3d$ $^4$D$_{1/2}$ & 104984.3238(50) & 104984.37 & -0.05\\
$3d$ $^4$D$_{3/2}$ & 104996.2343(50) & 104996.27 & -0.04\\
$3d$ $^4$D$_{5/2}$ & 105008.5141(50) & 105008.55 & -0.04\\
$3d$ $^2$D$_{3/2}$ & 105119.8418(49) & 105119.880 & -0.038\\
$3d$ $^2$D$_{5/2}$ & 105143.6799(50) & 105143.710 & -0.030\\
$5s$ $^2$P$_{1/2}$ & 110035.678(29) & 110035.720 & -0.042\\
$5s$ $^2$P$_{3/2}$ & 110103.793(36) & 110103.834 & -0.041\\
$4d$ $^2$F$_{5/2}$ & 110286.276(29) & 110286.305 & -0.029\\
$4d$ $^4$P$_{3/2}$ & 110299.937(29) & 110299.974 & -0.037\\
$4d$ $^2$F$_{7/2}$ & 110362.398(36) & 110362.462 & -0.064\\
$5d$ $^2$F$_{7/2}$ & 112891.180(36) & 112891.238 & -0.058\\
\hline
\end{tabular}
\end{table}

\subsection{Isotope shifts}

The finite mass $M$ of the nucleus results in a small nuclear motion in the center-of-mass reference frame, where the nuclear momentum $\mathbf{P}$ and electron momenta $\mathbf{p}_i$ are conserved: $\mathbf{P}=-\sum_{i}\mathbf{p}_i$.
The mass shift can be calculated from the expectation value of the nuclear kinetic energy operator
\begin{equation}
\frac{\mathbf{P}}{2M}=\frac{1}{2M}\sum_{i}\mathbf{p}^2_i + \frac{1}{2M}\sum_{i\neq j}\mathbf{p}_i\cdot\mathbf{p}_j .
\end{equation}
The first term on the right-hand side represents the Bohr shift or normal mass shift (NMS), while the second term refers to the specific mass shift (SMS).
The NMS is proportional to the atomic Rydberg constant and straightforwardly results in a blue shift for a heavier isotope.
The SMS is related to electron correlation, so that its magnitude and sign is highly dependent on the specific level involved.
Since the NMS and SMS terms depend quadratically on $\mathbf{p}_i$'s, they are often of the same order of magnitude, and in some cases are found to cancel~\cite{Eikema1994}.
We adopt the convention for the SMS such that
\begin{equation}
\mathrm{SMS}=-\left(\frac{1}{M_B} - \frac{1}{M_A}\right)\left(k_u - k_l\right)
\end{equation}
where isotopic masses follow $M_B>M_A$ and
\begin{equation}
k_{\{u,l\}}=\left<\psi_{\{u,l\}}\left|\sum_{i\neq j}\mathbf{p}_i\cdot\mathbf{p}_j \right|\psi_{\{u,l\}}\right>,
\end{equation}
which ensures that a positive SMS shifts in the same direction as the NMS.
In the following discussions we neglect the effects of nuclear field shifts and hyperfine structure~\cite{Cangiano1994,Jennerich2006} as these are smaller than the spectral resolution in the present study.

\subsection{Isotope shift in N I}

For the $2s^22p^3\,^4\mathrm{S}_{3/2} \rightarrow 2s^22p^23s\,{}^4\mathrm{P}_{J}$ transition in nitrogen
the normal mass shift, NMS($^{15}\mathrm{N}$) - NMS($^{14}\mathrm{N}$), amounts to 0.229~\wn,
while it is 0.242~\wn\ for the $2s^22p^3\,^4\mathrm{S}_{3/2} \rightarrow 2s2p^4\,{}^4\mathrm{P}_{J}$ transition.
As listed in Table~\ref{tab:ground N}, the isotopic shift $\Delta_{15-14}$ is not resolved for the transitions to the $2s^22p^23s\,{}^4\mathrm{P}_{J}$ levels while a clear isotope splitting is observed for the transitions to $2s2p^4\,{}^4\mathrm{P}_{J}$ levels shown in Fig.~\ref{N_2p4}.
The observed transitions originating from the metastable $2s^22p^3\,{}^2\mathrm{D}, \,{}^2\mathrm{P}$ states listed in Table~\ref{tab:meta N} do not display any discernible IS splittings.
The SMS for the $2s^22p^3\,^4\mathrm{S}_{3/2} \rightarrow 2s2p^4\,{}^4\mathrm{P}_{J}$ transitions is extracted at $+0.40(1)$~\wn.

A large IS is associated with the promotion of a $2s$ electron from the $2s^2 2p^3$ ground state to the $2s 2p^4$ configuration in the excited state. This is consistent with  Clark's ~\cite{Clark1983,Clark1984} calculations which showed that the dominant contributions to the $k$ integrals (and SMS) increase with the number of $2p$ electrons.
As a consequence when the number of $2p$ electrons in the upper state is larger than that in the lower state, SMS is positive and hence enhances the total IS in the $2s 2p^4$ excited state configuration.
On the other hand, when the number of $2p$ electrons in the upper state is less than that in the lower state SMS is negative and largely cancels the total IS for
transitions to the $2s^2 2p^2 nl$ states.
The same trend is found for transitions accessed by laser measurements in the infrared~\cite{Cangiano1994} and VUV range~\cite{Salumbides2005}.

\subsection{Isotope shift in C I}

For carbon, the $^{13}$C - $^{12}$C IS for the $2p^2$ $^3$P$_J'$ - $2p 3s$ $^3$P$_J$ transitions is about -0.10~\wn\ on average.
In comparison with Ref.~\cite{Haris2017}, there is an average difference of 0.009 \wn, hence smaller than the combined uncertainty.
Note that the IS presented in Ref.~\cite{Haris2017} are taken from theoretical values in Ref.~\cite{Berengut2006} with an estimated uncertainty of 0.004~\wn.
The IS determined in the present study is consistent with, but more accurate than the measurements of Haridass and Huber~\cite{Haridass1994}.
With an NMS of about 0.21~\wn\ for the C~I transitions, the SMS is derived to be -0.31(4)~\wn.
The negative SMS can be understood from the same arguments as given above, stemming from the smaller number of $2p$-electrons in upper state compared to lower state in the $^3$P$_J$ - $3s$ $^3$P$_J$ transitions.
On the other hand measurements involving core-changing transitions from the $2s^2 2p^2$ ground state to the $2s 2p^3$ excited configuration in C~I show a positive SMS resulting in large IS~\cite{Haridass1994,Labazan2005}.
These are consistent with the expected results from \emph{ab initio} calculations that employ different flavors of (post-)Hartree-Fock methods obtaining varying levels of accuracy~\cite{Clark1983,Berengut2006}.

\section{Conclusion}

Accurate measurements of transition energies in nitrogen and carbon atoms were obtained at an absolute accuracy of 0.025~\wn\ using VUV Fourier Transform spectroscopy with a synchrotron radiation source.
For $^{14}$N and $^{15}$N, transitions originating from the ground $2s^2 2p^3$ $^3$P states as well as from metastable states $2s^2 2p^3$ $^2$D and $2s 2p^3$ $^2$P states are observed.
For $^{12}$C and $^{13}$C, transition energies for $2s^22p^2$ $^3$P$_{J'}$ - $2s^22p 3s$ $^3$P$_J$ lines were measured.
The comprehensive dataset for N I is included in a reevaluation of the level energies of the excited states, in combination with data from a previous laser-based precision study~\cite{Salumbides2005}. This results in an averaged shift of -0.04 \wn\ with respect to the level energies reported in  the NIST database~\cite{NIST_ASD}.

The determination of isotope shifts for carbon and nitrogen in this study will be useful in assessing the effectiveness of various strategies in \emph{ab initio} calculations for many-electron atoms in particular towards the treatment of electron correlation.
Such tests on isotopic shifts will be complementary to benchmarking with absolute level energies, where the most accurate theoretical description for multi-electron remains to be a difficult challenge.

\section*{Acknowledgement}

The authors are grateful to the staff at the Soleil synchrotron for their hospitality and operation of the facility under project number 20190381.
WU acknowledges the European Research Council for an ERC-Advanced grant under the European Union's Horizon 2020 research and innovation programme (grant agreement No 670168).

%

\end{document}